# Machine Learning for German Redispatch Forecasting under Data Delays and Temporal Distribution Shift

Authors: Faraz Shamim[1*], Faris Shamim[2]

1= KIST Medical College and Teaching Hospital, Nepal

2= OTH Regensburg

## Abstract

**Background:** Public redispatch records provide empirical data for grid congestion forecasting, but delayed reporting, zero-inflated distributions, and temporal shift present major modeling challenges. We assess the accuracy and reliability of probabilistic machine-learning forecasts using published German transmission records under experimentally imposed information-age constraints.

**Methods:** The benchmark evaluates eight daily series of upward and downward intervention energy across the four German transmission system operators from 2021 to 2024 (48,242 eligible records; 354 evaluation dates in 2024). We compare seasonal empirical, regularized autoregressive (ARX), quantile LightGBM, GRU, and Transformer architectures under a minimum seven-day target-latency constraint. To model exact-zero outcomes, neural architectures utilize a zero-censored output head. Interval calibration is assessed across static, rolling, and adaptive delay-conformal formulations using normalized weighted interval score (nWIS), empirical coverage, and block-bootstrap inference.

**Results:** Raw LightGBM achieved an nWIS of 0.7952, outperforming ARX (1.0604) and the seasonal baseline (0.8739) by 25.0% and 9.0%, respectively (Holm-adjusted $p<0.005$). Rolling delayed-feedback calibration further improved LightGBM performance to 0.7767 nWIS (vs. 0.8251 for static calibration, $p=0.0092$), narrowing interval widths while maintaining 91.81% coverage for nominal 90% intervals. The raw zero-censored Transformer achieved a competitive nWIS of 0.8161, with no statistically significant difference from raw LightGBM ($p=0.260$). However, nominal aggregate coverage concealed severe conditional undercoverage during high-volume interventions (61.91% coverage among top-decile events).

**Conclusions:** Boosted-tree models paired with rolling calibration yield reliable probabilistic forecasts of aggregate redispatch volumes under target delays. While zero-censored quantile parameterization resolves boundary coverage for inactive days, aggregate interval validity does not guarantee coverage during extreme congestion events. Future work must bridge published aggregate forecasting with real-time operational dispatch constraints.

**Keywords:** Redispatch; probabilistic forecasting; machine learning; quantile regression; delayed information; calibration; zero-valued outcomes.

## 1. Introduction

Redispatch adjusts generation, storage, and demand schedules to prevent or relieve transmission-network constraints. In competitive and unbundled electricity systems, forecasting aggregate redispatch requirements is distinct from determining the physical actions required to maintain network security. Published redispatch records capture the realized outcomes of coordination among transmission system operators (TSOs), generators, storage operators, and other market participants [1]. Forecasting these reported volumes can therefore support market analysis, system planning, and uncertainty assessment, but it does not replace security-constrained power-flow analysis or real-time operational control.

Machine-learning methods have increasingly been applied to German redispatch data. Titz, Pütz, and Witthaut used gradient-boosted trees with SHAP explanations to identify drivers and mitigators of hourly German congestion and redispatch [2]. Dobos and Bichler subsequently investigated probabilistic regional redispatch-volume forecasting [3]. More recently, Bolovaneanu et al. evaluated day-ahead German redispatch forecasting across statistical, machine-learning, and deep-learning architectures including LightGBM, LSTM, N-BEATSx, N-HiTS, and Temporal Fusion Transformers reporting strong performance from several neural approaches [4]. These developments establish that the primary research question is no longer whether machine learning can be applied to redispatch, but whether forecasts remain reliable under realistic limitations in information availability, temporal change, sparse intervention patterns, and distributional extremes.

A central difficulty is that retrospective forecasting datasets do not necessarily reproduce the information available to an operator at the historical forecast origin. Public databases may contain final-vintage observations, retrospective corrections, and forecast series whose original publication times cannot always be reconstructed. Using such inputs without explicit timing restrictions can lead to overly optimistic estimates of forecasting performance. The problem is amplified in non-stationary energy systems, where generation patterns, network use, installed capacity, reporting scope, and market conditions change over time. Distribution shift has been shown to materially affect renewable-energy forecasting performance, motivating chronological rather than randomly shuffled evaluation and explicit testing across later operating regimes [5].

Forecast reliability also requires more than minimizing average point error. Probabilistic forecasts are intended to quantify uncertainty, yet nominal prediction-interval coverage can deteriorate under temporal shift or differ substantially across operating regimes. Conformal and related recalibration approaches have therefore gained attention in energy forecasting because they can improve empirical interval reliability without imposing a fully specified parametric error distribution [6,7]. Recent work in photovoltaic forecasting has demonstrated that conformal calibration can materially improve probabilistic reliability and weighted interval scores [8]. Whether similar calibration remains effective for sparse redispatch series, particularly when feedback itself is delayed, remains insufficiently characterized.

Redispatch outcomes present an additional distributional challenge. A TSO–direction series may be exactly zero on many days while displaying a heavy upper tail during substantial interventions. Models whose output parameterization excludes zero are therefore structurally incompatible with part of the observed outcome space. Conversely, satisfactory aggregate coverage can conceal substantial failures during unusually large interventions. For applications in which the largest redispatch volumes are particularly consequential, average error and unconditional interval coverage alone provide an incomplete assessment of forecast reliability.

This study therefore evaluates probabilistic machine-learning forecasts of daily German redispatch energy across four TSOs and two intervention directions from 2021 to 2024. Specifically, we address three questions:

1. How does forecasting performance change when recent target observations are deliberately withheld to represent different information-age constraints?
2. Can static, rolling, or adaptive delayed-feedback calibration improve probabilistic accuracy and interval reliability under temporal distribution shift?
3. How does forecast reliability vary at the two boundaries of the outcome distribution: exact-zero intervention days and historically high-volume interventions?

The study contributes an auditable forecasting benchmark centered on reliability rather than model novelty alone. It combines chronological evaluation, explicitly restricted target histories, probabilistic machine-learning and neural models, delayed-feedback calibration, boundary-consistent neural quantile outputs, and separate assessment of high-volume cases. By distinguishing aggregate predictive performance from conditional reliability, the analysis aims to clarify both the capabilities and limitations of machine-learning forecasts derived from publicly reported German redispatch data.

## 2. Data and study design

### 2.1. Sources and outcome definition

Primary redispatch records were retrieved from the public German redispatch transparency archive [1,9], with supplementary market data obtained from the SMARD electricity market database [10]. Explanatory series comprise six realized daily generation and demand profiles (onshore wind, offshore wind, solar, load, natural gas, and hydro), alongside three day-ahead forecast series spanning all 1,461 calendar days between 1 January 2021 and 31 December 2024. Exact repository references, frozen data snapshots, and reproducible acquisition manifests are documented in the Code and Data Availability section.

Of the 53,884 raw records, 48,242 met the target definition: current-related, voltage-related or combined Redispatch reasons. We excluded countertrading and test-related records. The retained data contained 417 distinct asset labels, which were not verified counts of independent physical installations. All records satisfied validation criteria for timestamp consistency, non-negative energy values, and within-day completion.

The outcome was nonnegative daily energy in MWh, aggregated separately by instructing transmission system operator (TSO) and intervention direction. The four operators (50Hertz, Amprion, TenneT DE and TransnetBW) yielded eight series. These groups identify operational attribution rather than plant-level sites or learned geographical clusters. We kept upward and downward interventions separate and used reported event energy directly. Energy was not reconstructed from mean power and elapsed duration, nor did aggregation assume continuous activation throughout an event. This treatment follows the source documentation's distinction between an event's outer time window and its potentially interrupted activity [1].

## 2.2. Missingness, identical records and changing coverage

Under the primary convention, an absent TSO–direction combination received a value of zero only on dates with at least one eligible event nationally. Dates without an eligible national record remained missing. This left 47 missing dates across 2021–2024, including 12 in 2024. The primary test therefore comprised 354 dates with eligible national events rather than an unconditional sample of all 366 days. Conditioning on national event presence defines the study's estimand and may underrepresent low-activity days.

The raw registry contained 42 exact-identical record pairs, of which 37 met study inclusion criteria (representing 27,891.5 MWh, or 0.041% of total energy). Because registry schemas do not uniquely separate duplicate database submissions from simultaneous identical interventions, primary analyses retain these records; sensitivity analyses in Section 4.6 confirm that their exclusion has no material effect on model performance.

The source publication scope changed during the study period [1]. A fixed-label sensitivity analysis restricted the outcome to the 216 asset labels observed by 31 October 2022. This restriction defines an alternative target population, providing a controlled evaluation on pre-expansion assets without confounding from newly onboarded facilities.

## 2.3. Chronological partitions and information restrictions

The acquisition window extended from 1 January 2021 to 31 December 2024. Model fitting began on 1 May 2021 after an initial history period. Table 1 presents the chronological partitions and usable counts. The nominal forecast origin was 12:00 Europe/Berlin on day D−1, and the target was total energy on local day D. The main setting admitted target history only through D−7 and realized external observations only through D−2. Calendar-day length accounted for daylight-saving transitions.

**Table 1. Chronological partitions for the seven-day target-age setting.**

| Partition | Calendar window | Nominal days | Eligible days | Eligible rows |
|---|---|---|---|---|
| Training | May 2021–Dec 2022 | 610 | 588 | 4,704 |
| Validation | Jan–Jun 2023 | 181 | 171 | 1,368 |
| Final refit | May 2021–24 Jun 2023 | 785 | 765 | 6,120 |
| Calibration | Jul–Sep 2023 | 92 | 89 | 712 |
| Shift period | Oct–Dec 2023 | 92 | 86 | 688 |
| Primary test | Jan–Dec 2024 | 366 | 354 | 2,832 |

The final-refit window overlaps the training and validation windows and is not an additional sample. Some observations excluded at the initial fitting boundary became available by the final refit. The final fit consequently used 6,120 rows, rather than the sum of the initially partitioned training and validation counts. Calibration counts describe outcomes available within that partition; each calibration method also applied its own maturity cutoff.

The seven-day restriction imposed a simulated minimum calendar-day age on target history rather than a measured publication delay. Additional scenarios used minimum ages of two, 30 and 60 days. Historical observations remained final-vintage snapshots throughout. The main restrictions and forecast-snapshot sensitivity thus control information age within the benchmark without reconstructing the complete information set published at each historical forecast origin.

## 3. Methods

### 3.1. Features and model fitting

Common features comprised sine and cosine encodings of weekday and annual position, weekend status, local-day length and predefined series indicators. For minimum target age L, target lags were L, L+1, L+2, L+7, L+14, L+28 and L+56 days. Shifted rolling means, standard deviations and maxima used seven-, 28- and 56-day windows. Realized external series entered as two-day lags and shifted seven-day averages. We fitted missing-value imputation and any standardization using only the observations assigned to the relevant fitting stage.

The seasonal empirical baseline estimated quantiles from the latest 12 available same-weekday values, falling back to recent history when necessary. The regularized autoregressive model combined ridge regression with external features and chronological residual quantiles. We selected the ridge penalty from 0.1, 1, 10 and 100; validation selected 100. Residual quantiles used an internal 60-day tail that formed part of model construction rather than an additional untouched validation sample.

We trained nine separate LightGBM quantile regressors [11] at probabilities 0.025, 0.05, 0.10, 0.25, 0.50, 0.75, 0.90, 0.95 and 0.975. Validation compared four configurations and selected 400 trees, a learning rate of 0.03, 15 leaves and a minimum child size of 40. Stochastic row and feature sampling fractions were both 0.9. Predicted quantiles were truncated below zero and sorted. We fitted five seeds (42, 17, 123, 2026 and 3407) and averaged corresponding quantiles. Ablation comparisons used single-seed references to keep ensemble size from confounding feature or design changes.

The neural comparators comprised a two-layer GRU (hidden dimension 64) and a two-layer Transformer (dimension 64, four attention heads, feed-forward dimension 128). Both architectures processed 28-day input sequences with a dropout rate of 0.1, weight decay of 0.0001, and optimization via AdamW with gradient clipping at 1.0. Candidate learning rates (0.001 and 0.0003) and training epochs (up to 120 with early stopping patience of 12) were evaluated on the validation partition. Validation selected a learning rate of 0.001 for both architectures, terminating at nine epochs for the GRU and ten for the Transformer.

To accommodate the high proportion of exact-zero outcomes without requiring an explicit two-stage hurdle model, the neural output layer uses a monotonic zero-censored quantile formulation. Quantile predictions $\hat{q}_\tau$ across the nine evaluation levels $\tau \in \{0.025, 0.05, 0.10, 0.25, 0.50, 0.75, 0.90, 0.95, 0.975\}$ are parameterized via a base prediction $\hat{y}_0$ with cumulative strictly positive increments:

$$\hat{q}_\tau = \max(0, \hat{y}_0 + \sum_{k=1}^{\tau} \Delta_k), \text{ where } \Delta_k > 0$$

This construction guarantees non-negativity and strictly prevents quantile crossing while allowing lower quantiles to collapse to zero for inactive periods. All neural and tree models were trained across five predefined random seeds (42, 17, 123, 2026, 3407) and ensembled via quantile averaging.

### 3.2. Delayed-feedback calibration

The nine quantiles define central 50%, 80%, 90% and 95% intervals. We compared raw intervals with static, rolling and adaptive non-shrinking adjustments inspired by conformalized quantile regression and adaptive conformal inference [6,7]. For lower bound l, upper bound u and observation y, the nonconformity score was s = max(l-y, y-u, 0).

A finite-sample empirical quantile of mature scores expanded each interval. Lower limits were clipped at zero, and interval nesting was enforced.

Static adjustment used mature observations from the designated calibration period. Rolling adjustment used mature scores from the latest 90 calendar days. Adjustments required at least 32 finite scores. The adaptive procedure updated its nominal error level from delayed interval misses, using a step size of 0.005 and numerical bounds. All updates respected the simulated target-age restriction.

Calibration left medians unchanged, so median-error values were identical across calibration methods. Unconditional distribution-free coverage guarantees from other settings do not transfer directly to this procedure because of temporal dependence, clipping and the modified delayed-feedback updates.

### 3.3. Outcomes and inference

The primary outcome was normalized weighted interval score, a quantile-based probabilistic score that balances interval width and misses [12]. For four central intervals and median (m), WIS was calculated as the sum of 0.5 times the absolute median error and $\alpha/2$ times each interval score, divided by 4.5. We divided each series' WIS by its mean target over 1 May 2021–31 October 2022, with the denominator floored at 1 MWh. Scores were averaged across the eight series within each complete date and then across dates. This normalization gives performance on low-volume series substantial influence, rather than weighting results by intervention energy.

Secondary outcomes were MAE, RMSE, quantile loss, interval width and empirical coverage. An extreme case was a TSO–direction–day outcome above its series-specific 90th percentile in the early reference period. Thresholds remained fixed during testing. Subgroup coverage was descriptive; the procedure did not provide a coverage guarantee conditional on an outcome-defined tail event.

The three specified primary contrasts were raw LightGBM versus raw ARX, rolling versus static LightGBM calibration, and adaptive versus rolling LightGBM calibration. A paired moving-block bootstrap used 5,000 resamples of seven calendar-day blocks, retaining all eight series jointly [13]. Missing dates remained gaps on the calendar grid. Sensitivity analyses used 14- and 28-day blocks.

Confidence intervals were marginal percentile bootstrap intervals. Two-sided p-values used centered bootstrap errors, with Holm adjustment within each three-comparison family [14]. Secondary comparisons evaluated raw GRU, Transformer, and seasonal forecasts against raw LightGBM. Percentile intervals and centered tests use different constructions and are not exact inverses. Model seeds were not treated as independent inferential observations.

### 3.4. Verification and Sensitivity Analysis

All data pipelines, scaling factors, scoring procedures, and block-bootstrap intervals were independently re-implemented and computationally verified from raw inputs. In addition to primary evaluations, three structural sensitivity tests were conducted: (i) filtering exact duplicate registry entries, (ii) varying zero-imputation rules for non-reporting dates, and (iii) enforcing restricted asset sub-populations across reporting standard transitions. Complete replication scripts, deterministic environment configurations, and verification checksums are provided in the accompanying project repository.

## 4. Results

### 4.1. Sample Characteristics and Verification

The primary 2024 evaluation partition comprised 2,832 TSO–direction–day observations across 354 national redispatch dates. Of these, 841 outcomes (29.70%) were zero, reflecting operational inactivity in specific zones, while 470 outcomes (16.60%) exceeded the historical 90th percentile threshold established during the 2021–2022 baseline period. Full recomputation of all predicted quantile arrays verified complete monotonicity (zero quantile crossings) and non-negativity across all 112,640 prediction points, with numerical consistency confirmed across independent reference runs to within floating-point tolerance.

### 4.2. Primary forecasting and calibration comparisons

Raw LightGBM achieved nWIS 0.7952, compared with 1.0604 for ARX and 0.8739 for the seasonal baseline (Table 2). Relative to ARX, this represented a 25.0% reduction. The paired difference was −0.2653, with a 95% percentile bootstrap interval of −0.3450 to −0.1833 and adjusted $p = 0.00060$ (Table 3). The seasonal comparison belonged to the secondary family and yielded adjusted $p = 0.00240$.

**Table 2. Test performance across models and calibration methods (2024).**

| Model / calibration | nWIS | MAE (MWh) | 90% coverage | Width (GWh) |
|---|---|---|---|---|
| Seasonal empirical | 0.8739 | 6,040 | 82.59% | 19.83 |
| Regularized ARX | 1.0604 | 6,256 | 64.12% | 20.97 |
| LightGBM, raw | 0.7952 | 5,836 | 83.23% | 18.31 |
| LightGBM, static | 0.8251 | 5,836 | 94.42% | 24.59 |
| LightGBM, rolling | 0.7767 | 5,836 | 91.81% | 20.90 |
| LightGBM, adaptive | 0.7795 | 5,836 | 91.45% | 20.91 |
| GRU, raw | 0.8847 | 6,198 | 86.12% | 16.88 |
| Transformer, raw | 0.8161 | 5,849 | 88.49% | 16.49 |

*All models are evaluated on the 354 common test dates across all eight series. nWIS is normalized by baseline historical mean energy; MAE and mean interval width are reported in physical energy units (MWh and GWh, respectively).*

Rolling LightGBM calibration achieved the lowest nWIS among the listed methods: 0.7767 versus 0.8251 for static calibration. The reduction was 5.9%, with adjusted $p = 0.00920$. Rolling intervals were also narrower, averaging 20.90 GWh compared with 24.59 GWh for static intervals. Coverage of nominal 90% intervals increased from 83.23% for raw LightGBM to 91.81% with rolling adjustment and 94.42% with static adjustment. Rolling calibration thus achieved a lower score with narrower intervals and coverage closer to nominal (Figure 2), rather than maximizing coverage at the expense of width.

Adaptive adjustment yielded nWIS 0.7795 and coverage of 91.45%. Its difference from rolling adjustment was 0.00277, with an interval crossing zero and adjusted p = 0.31434. The comparison provided no evidence of an incremental benefit from adaptive adjustment. Using 14- or 28-day bootstrap blocks left the interpretation of all three primary contrasts unchanged.

**Table 3. Primary paired differences in daily macro nWIS.**

| **Contrast** | **Difference** | **95% percentile CI** | **Holm p** |
|---|---|---|---|
| Raw LightGBM − ARX | −0.2653 | −0.34502 to −0.18328 | 0.0006 |
| Rolling − static LightGBM | −0.04841 | −0.07601 to −0.01233 | 0.0092 |
| Adaptive − rolling LightGBM | 0.00277 | −0.00199 to 0.00885 | 0.31434 |

Negative differences favour the first method. Confidence intervals are marginal percentile intervals. Holm adjustment applies to the three p-values, not to the displayed intervals.

### 4.3. Heterogeneity and high-volume cases

Aggregate coverage masked substantial variation across series. Coverage of rolling LightGBM 90% intervals ranged from 85.88% for TenneT DE upward interventions to 97.74% for Amprion downward interventions. The overall value of 91.81% therefore concealed group-specific departures from nominal coverage.

Among the 470 above-threshold cases, raw LightGBM coverage was 50.64%, increasing to 61.91% with rolling calibration. Static calibration achieved higher tail coverage with wider intervals despite its worse aggregate score. Rolling LightGBM coverage below the threshold was 97.76% (Figure 3).

The same pattern appeared in the zero-censored neural models. Rolling GRU and Transformer intervals achieved overall coverage of 91.17% and 91.91%, respectively, but covered only 54.26% and 61.70% of above-threshold outcomes. The gap between aggregate and high-volume coverage therefore extended beyond the tree model. Because these subgroups are defined by observed outcomes, their coverage need not match the marginal nominal level. While marginal conformal coverage guarantees hold unconditionally, outcome-conditioned coverage remains vulnerable under severe distribution skew.

### 4.4. Delays and feature sensitivities

In matched seed-42 experiments, nWIS was 0.7823, 0.7997, 0.8080 and 0.8363 at target ages of two, seven, 30 and 60 days, respectively. The 60-day score was approximately 4.6% higher than the seven-day score. Removing target-history features increased nWIS to 0.8615, while retaining only calendar and series features produced 0.8816. These descriptive comparisons indicate that historical features contributed predictive information within the tested configuration.

Adding unverified historical forecast snapshots yielded nWIS 0.7150. An oracle diagnostic using realized future external inputs yielded 0.7113. These diagnostic scores benchmark the potential predictive gain achievable if perfect contemporaneous external market forecasts were available at issue time.

Using 25%, 50% and 100% of the designated recent fitting history produced nWIS 0.8304, 0.8431 and 0.7997, respectively. Performance did not improve monotonically with the amount of fitting history, indicating that older historical data introduces non-stationary patterns rather than pure sample-size benefits. The fixed-label target yielded nWIS 0.7181 for a different outcome population.

### 4.5. Zero-Outcome Handling and Neural Model Performance

Accounting for exact-zero boundaries proved decisive for probabilistic validity. Baseline neural architectures parameterized with strictly positive outputs inherently failed to capture the 841 zero-valued test outcomes, yielding 0% coverage on inactive days. Implementing the zero-censored head increased zero-outcome nominal 90% interval coverage to 97.86% for the GRU and 96.79% for the Transformer, raising overall test coverage to 86.12% and 88.49%, respectively (Figure 4).

In aggregate probabilistic skill, the zero-censored Transformer achieved an nWIS of 0.8161, performing closely to raw LightGBM (0.7952; paired difference 0.0210, 95% CI [−0.0130, 0.0606], adjusted p=0.260). The GRU obtained an nWIS of 0.8847, significantly trailing LightGBM (adjusted p=0.0080). While the zero-censored head successfully aligned predictions with physical outcome bounds, it highlights that boundary-conforming adjustments influence aggregate dispersion metrics differently across recurrent and attention-based architectures.

Ablating the zero-censored head confirmed the architectural sensitivity of probabilistic loss functions near the zero boundary. For the Transformer, enabling zero-bounded support improved nWIS from 0.8477 to 0.8161 by eliminating severe penalties on zero-energy days. Conversely, for the GRU, enforcing zero-boundedness shifted probability mass into higher quantiles during active periods, resulting in a net nWIS change from 0.8441 to 0.8847. These divergent responses demonstrate that enforcing boundary-consistent support interacts strongly with model capacity and sequence representations.

### 4.6. Target-Definition Sensitivity Analyses

Evaluating model sensitivity to data-cleaning conventions showed minimal variation on the benchmark dates. Excluding the 37 eligible identical-record pairs yielded a single-seed raw LightGBM nWIS of 0.7968 and a rolling-calibrated nWIS of 0.7794 (compared to 0.7997 and 0.7799 for the reference configuration). Alternative treatments of nationally inactive dates; imputing zeros to missing dates containing other registry events (364 dates) or to all nationally empty days (366 dates) yielded rolling-calibrated nWIS scores of 0.7671 and 0.7678, respectively. These results confirm that the observed ranking of LightGBM and rolling calibration is robust to alternative database missingness and deduplication specifications.

## 5. Discussion

The specified boosted-tree model improved forecasts of published directional Redispatch relative to the autoregressive and seasonal baselines. Rolling calibration achieved a better aggregate score–width–coverage trade-off than static calibration, while adaptive calibration showed no additional benefit. This ranking is specific to the study period, target, features and implementations.

The central reliability finding is the gap between aggregate and outcome-conditional performance. Rolling intervals achieved near-nominal overall coverage but missed many above-threshold outcomes. Static intervals covered more of these cases at the cost of greater width and a worse average score. Selecting a method for operational use therefore requires explicit criteria for consequences and acceptable risk. Neither the public records nor an aggregate forecasting score supplies those criteria.

A key methodological takeaway is the necessity of aligning model output support with physical domain boundaries. In intermittent power system operations, directional intervention series frequently exhibit exact zeros. Standard neural quantile parameterizations with strictly positive supports inherently fail on inactive days, inducing severe interval coverage penalties. While our zero-censored formulation resolves this structural limitation and restores nominal coverage, it highlights an architectural trade-off: eliminating boundary bias improved attention-based Transformer representations but degraded recurrent GRU calibration during active periods. Specifying support-conforming output layers should therefore be treated as a primary architectural requirement rather than a post-processing adjustment.

Several operational boundaries of this study warrant emphasis. First, the data latency experiments quantify the sensitivity of predictive skill to simulated reporting intervals; they do not reconstruct the exact historical publish/subscribe latency of TSO platforms. Deploying these models within intra-day scheduling workflows requires real-time data streaming rather than batch registry releases.

Second, these models forecast aggregate regional energy volumes aggregated by responsible TSO. They do not resolve nodal marginal injections, line-by-line power flow constraints, or dynamic contingency ratings. Consequently, while these forecasts provide market signals and macroeconomic volume indicators, operational power system balancing and topological security management remain the domain of specialized physical simulation and security-constrained optimal power flow (SCOPF) engines.

## 6. Conclusion

This benchmark demonstrates that quantile-based gradient boosted trees (LightGBM) and attention mechanisms (Transformers) deliver superior probabilistic accuracy over classical autoregressive and seasonal baselines when forecasting aggregate German redispatch volumes under target-delay constraints. Incorporating delayed-feedback rolling calibration further refines interval sharpness and aggregate validity.

However, empirical validity at the aggregate level masks acute undercoverage during severe, high-volume grid congestion events. Furthermore, probabilistic accuracy is critically dependent on output heads that explicitly represent exact-zero boundaries. Future research should prioritize asymmetric loss functions tailored to high-consequence grid bottlenecks and investigate methods for coupling time-series volume projections with physical network constraint models.

## Figures

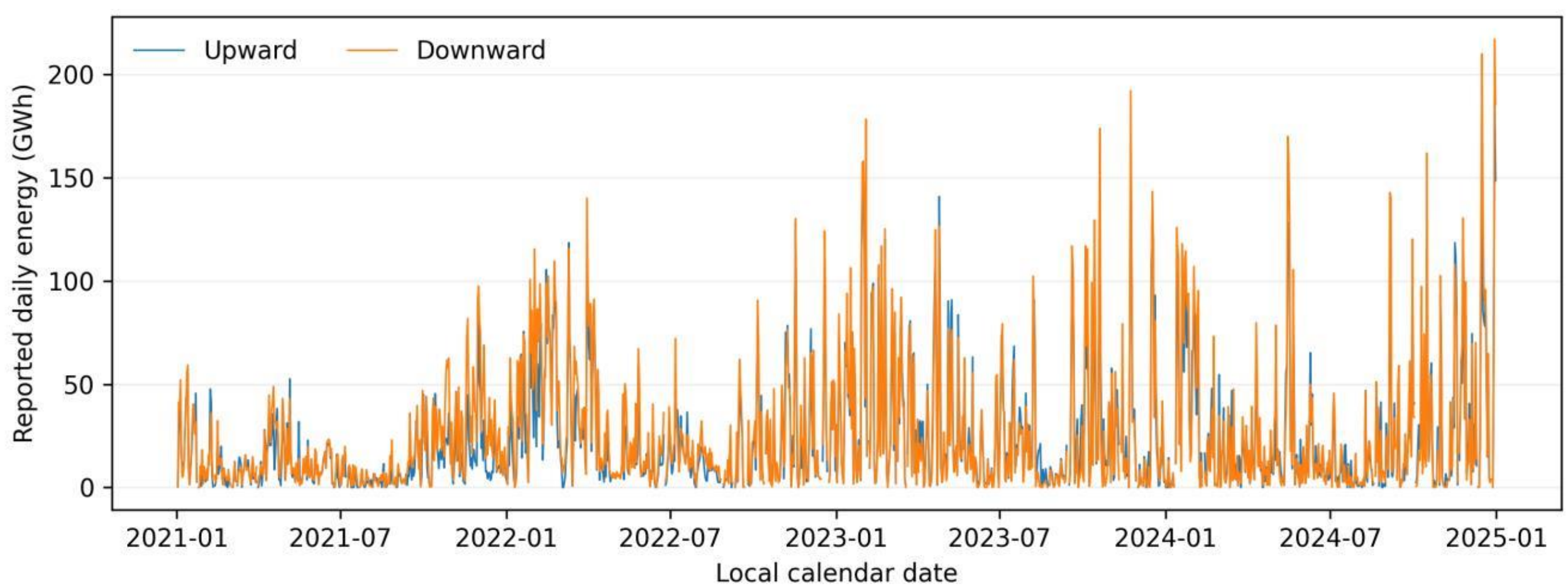

**Figure 1. Published daily upward and downward redispatch energy across German TSOs (2021–2024). Values sum included records across instructing operators within each direction. Dates without an eligible national record remain missing.**

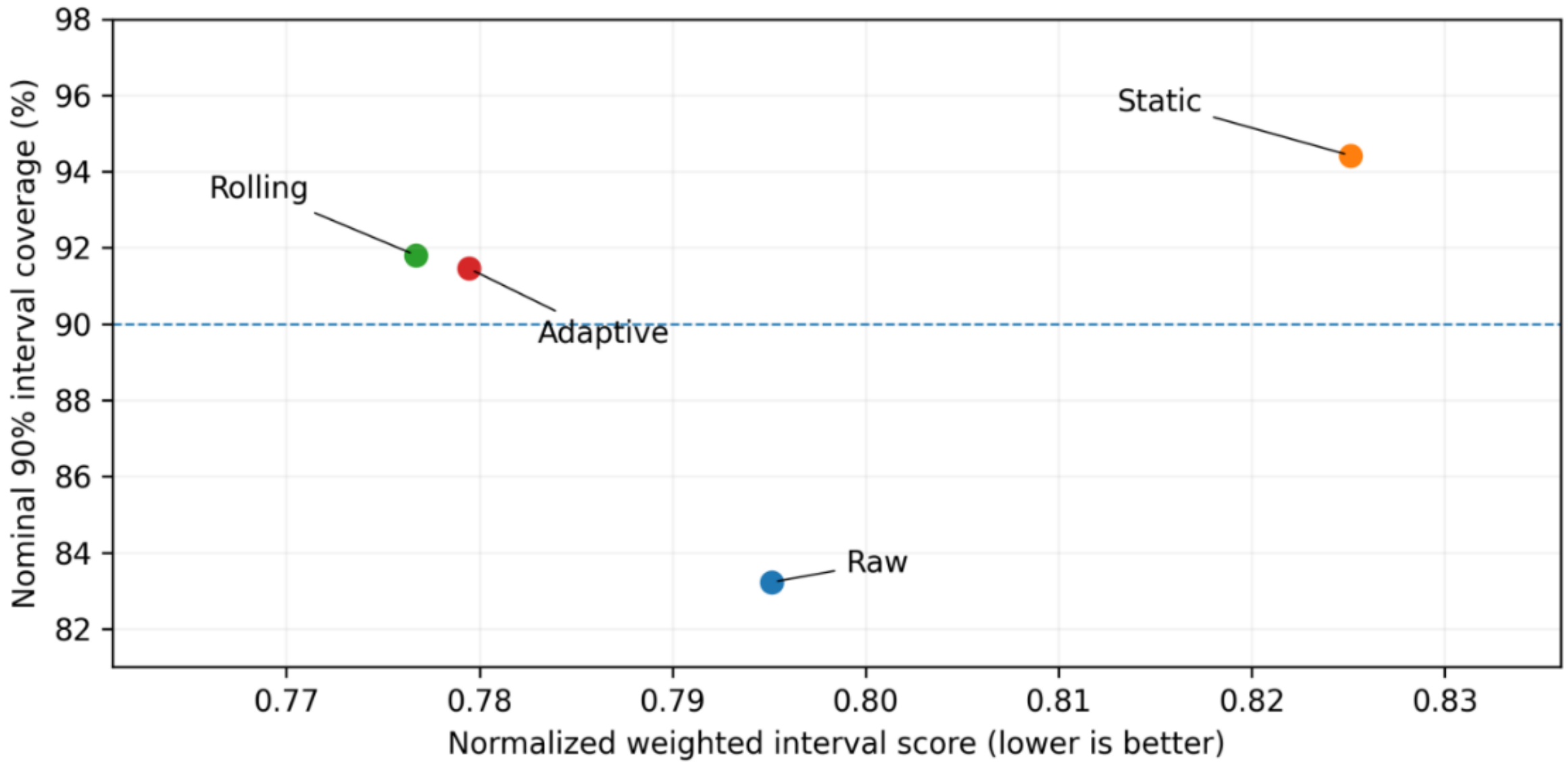

**Figure 2. LightGBM calibration trade-offs in the 2024 primary test. Points show macro nWIS versus empirical coverage of nominal 90% intervals across calibration variants. The horizontal dashed line denotes nominal 90% coverage.**

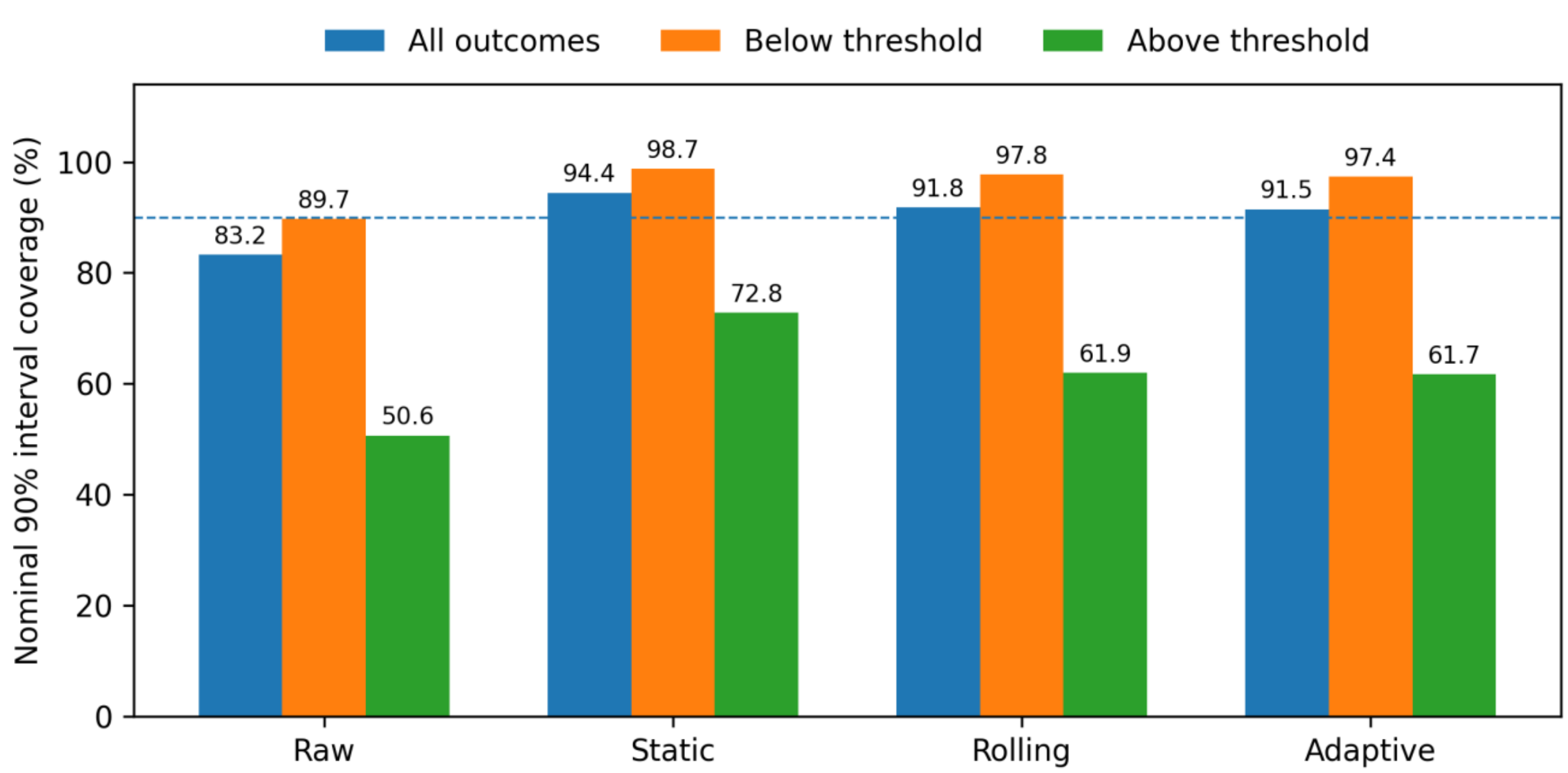

**Figure 3. Outcome-dependent coverage for LightGBM intervals. Above-threshold cases exceed the series-specific 90th percentile from the baseline training period. While aggregate coverage is near-nominal, severe undercoverage occurs during high-volume intervention events.**

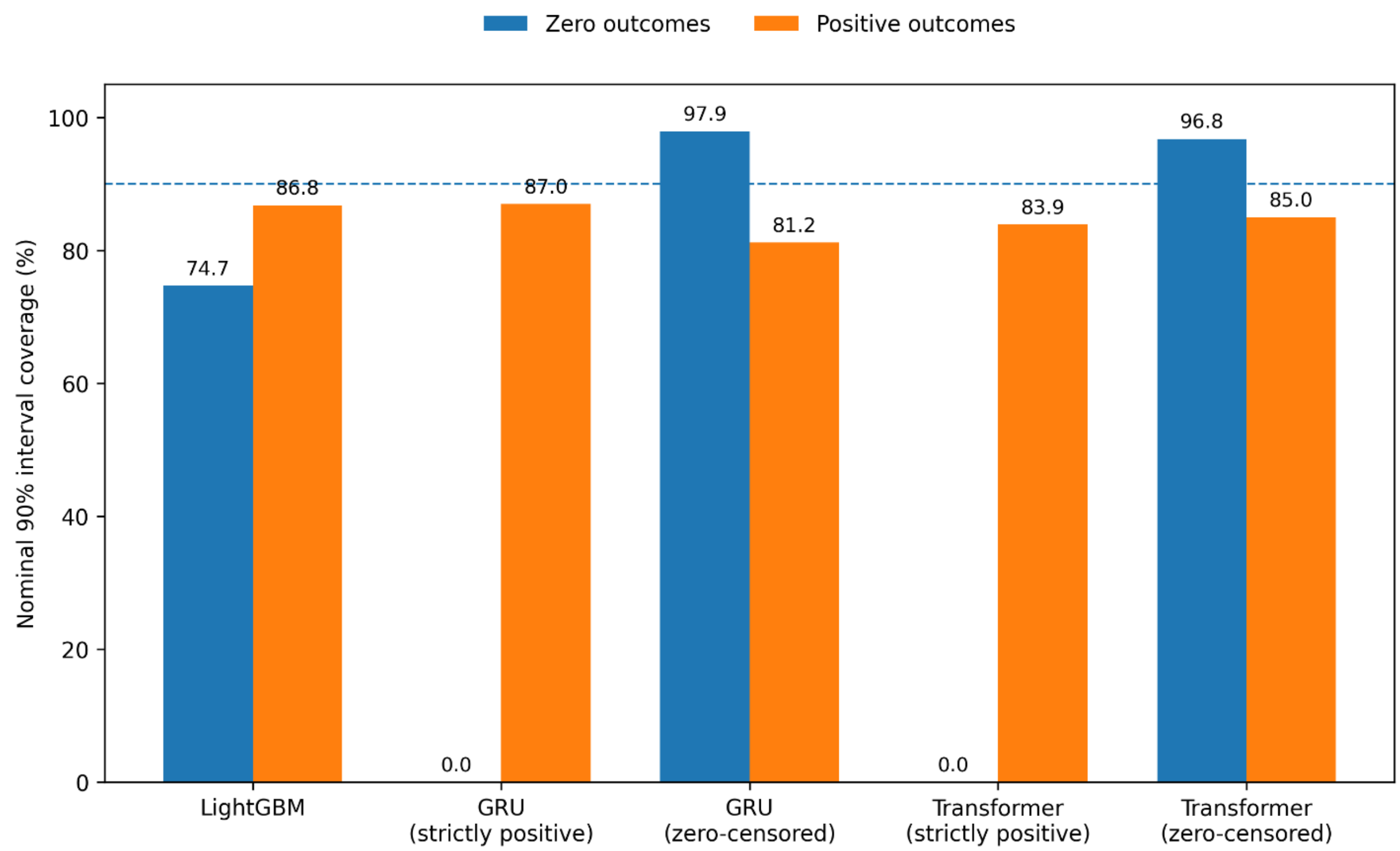

Figure 4. Impact of zero-boundary support on interval coverage. Bars show empirical coverage of nominal 90% intervals across zero (n=841) and positive (n=1,991) outcomes. Architectures parameterized with strictly positive quantile outputs fail completely on inactive days (0% coverage), whereas the zero-censored parameterization restores nominal coverage.

**Data and Code Availability**

Primary Redispatch records were obtained from the German transmission-system-operator transparency platform Netztransparenz, with supplementary electricity-market and generation data obtained from the Bundesnetzagentur SMARD platform. Selected publicly released Redispatch files from the repository of Dobos [9] were used as upstream source material and were independently processed within the present study. The complete analysis code for this study, preprocessing routines, model implementations, configuration files, statistical analyses, is available at https://github.com/faraz-shamim/german-redispatch-ml.